**NoSQL security: can my data-driven decision-making be affected from outside?**

**Anastasija Nikiforova[1,2], Artjoms Daskevics[3], Otmane Azeroual[4]**

**1** University of Tartu, Faculty of Science and Technology, Institute of Computer Science, Chair of Software Engineering, Narva mnt 18, Tartu, Estonia, Nikiforova.Anastasija@gmail.com, ORCID: 0000-0002-0532-3488

**2** European Open Science Cloud, 1050 Brussels, Belgium

**3** University of Latvia, Raina Boulevard 19, LV-1050 Riga, Latvia, artjoms.daskevics@gmail.com

**4** German Centre for Higher Education Research and Science Studies (DZHW), Schützenstraße 6A, 10117, Berlin, Germany, azeroual@dzhw.eu, ORCID: 0000-0002-5225-389X

**Abstract:** Nowadays, there are billions interconnected devices forming Cyber-Physical Systems, Internet of Things (IoT) and Industrial Internet of Things (IIoT) ecosystems. With an increasing number of devices and systems in use, amount and the value of data, the risks of security breaches increase. One of these risks is posed by open data sources, by which are meant databases, which are not properly protected. These poorly protected databases are accessible to external actors, which poses a serious risk to the data holder and the results of data-related activities such as analysis, forecasting, monitoring, decision-making, policy development, and the whole contemporary society. This chapter aims at examining the state of the security of open data databases representing both relational databases and NoSQL, with a particular focus on a later category.

**Introduction**

Today, with the technological advances affecting economic and socio-political landscapes, we deal with - create, generate, store, process, interpret and analyse – huge amounts of data on a daily basis in both private and public sectors (Visvizi & Lytras, 2019). They can be both structured, semi- structured and unstructured. Here, the concept of big data comes. These data, especially big data, form the basis for data-centric decision-making (Cui et al., 2014). This is all the more so in times of crisis, such as COVID-19 pandemics, since crisis management is typically a data-driven process of setting strategies, developing and coordinating actions and fiscal-, human-, time- resource allocation. This means that the data forms the basis for forecasting, predicting, mapping, tracking, monitoring, and raising awareness about the events (Corsi et al., 2021), contributing to decision-making and policy development as well as the whole contemporary society. They have both direct and indirect values, which means that corrupted or modified data may affect the results of the actions carried out with them and lead to inaccurate or even completely incorrect results and decisions. If used as input, they can make even very advanced algorithms relatively useless.

Therefore, both database management and the security of data and data storage in particular are very important for both technicians and non-technicians. Unfortunately, NoSQL databases are often found to be weakly protected and secured, which leads to many breaches and data leakages (Goel et al., 2021; Tran et al., 2019; Derbeko et al., 2016; Sahafizadeh et al., 2015; Daskevics et al., 2021). However, nevertheless many companies like Google and Amazon have abandoned traditional database technology in favour of proprietary data stores called key-value stores (Chang et al. 2006; DeCandia et al. 2007).

As regards the latter, i.e. data storages, one of the risks is posed by open data sources – databases, which are not properly protected, therefore they are available and accessible to external actors outside the organization. Although it may be surprising, the number of such databases is enormous. In many cases this is caused by misconfiguration, where the responsibility falls to database holders, in other cases there are vulnerabilities in the products, where apart of proper configuration additional security mechanisms are needed. Naturally, a series of questions arise: how to find out whether your database is visible and even accessible outside the organization? What information may be gathered from it? And whether stronger security mechanisms are needed or is the vulnerability rather related to internal configuration or the database in use? Although some questions may be partly answered by referring to Common Vulnerability and Exposure (CVE) Details and other sources summarizing vulnerabilities and patches on different services, this information may be too general. Therefore, penetration testing could be the answer.

The objective of this chapter is to provide an insight into the security of different databases, with a focus on NoSQL databases. First, the importance of this topic will be motivated by providing insight into data leaks by referring to some real-world examples (Section I). Section II will provide a brief overview of the causes of these data leaks and some others database security concerns. Section III will provide an insight on techniques and methods used to assess and maintain the level of the data storage security in use. It will cover both very general and rather static, and dynamic solutions. It will give an insight into current knowledge about the security of some NoSQL databases. Then the Search Engines for the Internet of Everything- (IoE) also known as Internet of Things Search Engines- (IoTSE-) based tool called ShoBEVODSDT, which has been recently developed, will be presented in Section IV. Section V will present the results obtained by applying ShoBEVODSDT to eight types of data storage facilities. Then, the discussion on the current security state of these databases and potential bottlenecks that need to be considered when any of them is being used, or a choice should be made in favour of any of them, will be provided.

**Data leakages**

The database is part of each system that is becoming even more popular as a result of digital transformations and Industry 4.0 reality (Kumar et al. 2019). With the increase of the usage of data and databases, particularly of NoSQL databases sometimes found to be most appropriate in the context of Big Data, IoT and IIoT, the topicality of database security increases as well. Unfortunately, NoSQL databases are mostly characterized by weak security, which leads to many cases of data leaks (Ferrari et al. 2020). Let us provide some evidences of this, i.e. provide some examples of data leaks.

According to one of the most up-to-date lists on major security breaches compiled by Tunggal and UpGuard (Tunggal, 2021), a large number of data leaks occur due to unsecured databases. Let's look at some of these breaches happened in the last 5 years:

1. ZOOM data breach occurred in 2020 lead to the sale or free publication of 500 000 accounts' data in a dark web;
2. Yahoo data breach in 2017 allowed hackers to compromise 3 billion user accounts. The investigation found that users' passwords in clear text, payment card data and bank information were not stolen. Yet it remains one of the largest data breaches in history;
3. Aadhaar data breach in 2018 resulted in the leakage of more than a billion Indian citizens' personal information stored in the world's largest biometric database, which could be purchased online, including data on names, their unique 12-digit identity numbers and bank details;
4. LinkedIn data breach in 2021, when 700 million LinkedIn users' data was brought for sale in a dark web. That exposure affected 92% of the total users, i.e. 756 million users;
5. Starwood (Marriott) data breach in 2018, when approximately 500 million hotel customers' data were stolen, when the attackers had gained an access to the Starwood system in 2014 and remained in the system after Marriott acquired Starwood in 2016. What is more, this discovery was not made until 2018;
6. Twitter data breach in 2018 and 330 million passwords exposed in an internal log, making all user passwords available to the internal network.
7. Uber data breach in 2016 and personal information of 57 million Uber users and 600 000 drivers exposed.

The data leaks listed above give some insight on how frequently and how many users data were exposed, the level of detail of which may vary from IDs and name to addresses and bank details. However, this information does not provide an insight on fiscal losses resulting from these data leaks. This is because, firstly, it is difficult to assess such losses (although such assessments tend to be published). Secondly, financial losses are not the only ones affecting the organisations concerned, since, in addition to direct losses, they tend to suffer indirectly, i.e. their reputation is affected negatively and can lead to losses of both clients / users and partners. In addition, although examples of users' data leakages have been covered here, it is clear that in many cases not only client / user data is the target of attackers and invaders, and other data make sense for attackers as well.

Perhaps the most provocative database in this respect is MongoDB, where 54 000 databases became accessible on the Internet in 2018, that lead to data leaks of 2.4 million of telemedicine vendor patients (Davis et al., 2018). According to Bizga (2020), it is a very common trend to target / compromise unprotected MongoDB servers for hackers, wiping out their data and leaving ransom notes outlining the threat of leaking stolen data and reporting to owners for GDPR violations. Thus in 2020 more than 22 000 ransom notes were left in the exposed MongoDB databases, i.e. nearly 47% of all MongoDB databases that are accessible online.

A similar trend is also observed when the security of MongoDB databases is assessed using the IoTSE and searching for open databases accessible via the Internet (questions like "how it works?" and "what are the results?" will be answered a bit later). Recent findings showed that the most popular IoTSE – Shodan – lists more than 15 000 affected MongoDB, while another popular IoTSE – Binary Edge – around 23 000. However, although these results are mainly discussed referring to MongoDB, it is not the only vulnerable or easily accessible database. According to other studies, including lists of the largest data leaks, ElasticSearch is also found highly vulnerable, sometimes followed by Cassandra (Panda, 2019; Tunggal, 2021; Ferrari et al., 2020).

However, it should be acknowledged that some improvements have been made in this respect in recent years, although this remains a problem. In addition, although the low security level of NoSQL databases is widely discussed, it cannot be stated that SQL databases' holders do not risktheir data leaking. Some evidence of this will be provided in the later sections. Moreover, given that MongoDB is not the only NoSQL database, more than the MongoDB will be referred to, covering some other NoSQL and SQL databases. This will be done using a tool based on the above mentioned IoTSEs that are described as relatively easy to use. This also explains their popularity and appropriateness for the purpose of this study.

**Security concerns**
Let us now provide a brief overview of what might be the causes of these data leaks and some others database security concerns, primarily referring to NoSQL databases. It is not a secret that there exist some indexes and registries to be used to identify the weakest areas of the service, such as the database, and find out the latest "threats" by which is characterized the service in use. Perhaps the most popular of them is Common Vulnerability and Exposure (CVE) Details (CVE Details, 2022). It provides an index of registered vulnerabilities of various services dividing them in 13 categories, including Denial of Service (DoS), Code Execution, SQL injection, HTTP response splitting, Gain information, Gain Privileges, CSRF and File Inclusion etc. However, although they are very valuable and useful, same as other registries such as VulDB and NVD – National Vulnerability Database, they are rather static and general and do not provide an information on the actual database in use, e.g. whether the database is visible and can be accessed from outside of the organization. Thus, additional steps should be taken inspecting the particular database used within the organisation. Before referring to this, another point to be elaborated on is related to database-specific security and privacy concerns, since the above listed categories used by the CVE are rather service-agnostic.

For this purpose, let us refer to Goel et al. (2021), who have conducted a systematic analysis ofprivacy-related literature in the context of NoSQL databases, using a pattern-based approach to identify what the most widely occurring "privacy-breaching" issues are. The authors refer to themas "privacy-breaching patterns". Goel et al. have managed to identify 6 patterns:

- **malicious query introduction** that occurs when a person with malicious intent interferes with the system and modifies a NoSQL query so that it can read or even modify a database or change data in a web application. Malicious queries allow users to manipulate the back-end of database by adding, modifying, or deleting data. They can be caused by either injection attacks or insider attacks. For injection attacks, perhaps the most commonly known are (a) tautologies injecting code in a conditional statement, (b) union queries used to bypass authentication and extract data, (c) JavaScript injections, which allows intruder to execute JavaScript to perform complex tasks, (d) piggy-backed queries that "exploits assumptions in the interpretation of escape sequences' special characters" to insert additional queries into theoriginal one, and (e) origin violation that uses HTTP REST APIs to access database fromanother domain. Insider attacks, in turn, refers to "the use of queries to gain unauthorised access to information";

- **accidental re-identification**, which refers to cases, where a person can be re-identifiedon the basis of the query output, despite the mechanisms in place to ensure data privacy.This, in turn, can happen by executing a complex query, resulting in a small output set, e.g. advanced queries or data processing capabilities such as MapReduce and an aggregation are used, that allows to re-identify artefacts such as individuals that raiseprivacy concerns. In other words, this may disclose / reveal sensitive data stored in thedatabase;

- **weak authentication** that refers to cases when NoSQL databases provide poor password storage mechanisms or are limited to no authentication capabilities. As an example, MongoDB and Redisdo not provide authentication by default, allowing an intruder to get an access to the system. It, in turn, can lead to "an individual with malicious intent gaining access to the system", which can lead to attacks such as masquerade and hijacking;

- **coarse-grained access control**, which is the process of "controlling access to data and resources in a system". It is usually done by associating different types of users with a certain set of rules based on their roles and responsibilities. NoSQL databases provide different levels of access control support, which typically can be divided in (a) role-based and (b) data-based;
- **vulnerable data in motion** that poses a risk to the privacy of data, if the cluster uses poor security mechanisms that can compromise operations of the clusters;
- **vulnerable data at rest** that refers to weak "data at rest" encryption or the use of datastorage mechanisms susceptible to external attacks.

According to the authors' experiment, weak authentication and vulnerable data at rest were identified as the most widespread threats in 8 NoSQL databases (MongoDB, CouchDB, Redis, Aerospike, Cassandra, Hbase, Neo4j, OrientDB), with weak authentication found to be the most significant among the assessed instances. Another study (Ferrari et al., 2020) investigated the use of the most popular NoSQL databases, focusing on misconfiguration analysis that can lead to security and privacy problems. The authors developed a tool that automatically scans IP subnets to identify exposed services, as well as performs security analyses. Their analysis showed that in 0.18% of IP addresses they have analysed, database in use is misconfigured. The risks associated with the services exposed range from data leakages that can pose a significant threat to users' privacy, to manipulation of resource data stored in vulnerable databases, which can pose a significant threat to the reputation of web services.

To sum up, although the nature of attacks may vary significantly, one of the main security concern is related to the fact that many NoSQL databases are less likely to provide security measures, including sometimes very primitive and simple measures such as an authentication, authorization (Sahafizadeh et al., 2015; Bada et al., 2015) and data encryption. They stand for the "open databases". As a result, according to recent observations, more than half of the known data leaks over the past years are leaks from open databases (Habr, 2022).

**Methods and techniques for inspecting the security of data storage facility**

There are many methods and techniques to check the overall security status of an artefact, such as a database. They mainly refer to vulnerability registries and indices / databases etc., as well as methods to assess the current state of the artefact in use, i.e. the local database used in the organisation. As for the approaches to testing the system in use, one of the most popular approaches is to perform an attack on a system such as an injection or CSRF (Cross-Site Request Forgery) as Ron et al. do (Ron et al., 2016). They demonstrate that while the emergence of new query formats makes old SQL injection techniques irrelevant, NoSQL databases are not immune to injections. To prevent breaches and data leaks, the authors recommend using Dynamic Application Security Testing (DAST) and static code analysis to find any injection vulnerabilities if coding guidelines were not followed. They stress that DAST should be seen as more credible and reliable, and allow a better understanding of the current state of the system and the need to improve security. Although they deal mainly with aspects not actually related to this study, the authors highlight the importance of a proper authentication mechanism and access control to avoid or at least reduce the risks of more advanced attacks they deal with. However, the source of these "more advanced" techniques, which often used to test the system used, is Open Web Application Security Project (OWASP), which collects examples of testing a specific product that meets the OWASP goal, i.e. to improve the security of software (OWASP Foundation, 2013).

For registries, perhaps the most popular index used for a variety of services is the above mentioned CVE Details. CVE Details provide data not only on databases having a broader scope, however the number of databases covered by it is limited. More precisely, some popular databases, such as Memcahced, ElasticSearch, characterized by a high number of vulnerabilities and leaks in recentyears, are not covered by it. This registry can therefore be used as a complimentary source, but in many cases it will not be applicable. Not least popular is NVD – National Vulnerability Database – the U.S. government repository that includes databases of security checklist references, security-related software flaws, misconfigurations, product names, and impact metrics. But as it was above-mentioned, the nature of these sources is rather static and general. They provide information on the general database management system-related issues rather than on the database in use. Thus, the question on whether the database is visible and can be accessed from outside of the organization should be answered by database holders.

Here the blind and *double blind testing* (also called double blind penetration or pentesting) in particular comes. It is sometimes considered tobe one of the most objective testing methods. This objectivity is ensured by the absence of preparatory works for testing, which may have affect the test results, while the pentester is an external actor without prior knowledge of the system and its specificities. Blind pentesting requires specific tools supporting testing, i.e. tool to find basicinformation on a system under test (Daskevics and Nikiforova, 2021a). This stage is usually linked to the evaluation / assessment process that can be done for both the entire environment and each component forming it, where the assessment can be carried out by the internal or the external auditor (Ramadhan et al., 2020). An internal audit may be carried out in a shorter time, providing information on existing threats and previous attacks, while external data audits and sources may help to facilitate knowledge of events in a broader cyberspace (Samtani et al., 2020). During the assessment process, the external auditors are expected to obtain as much information as possible in relation to the target. Perhaps the most valuable supporter of this task is vulnerability scanner (Burns et al., 2007). In order to be eligible to conduct more comprehensive analysis, it should conduct non-intrusive testing, mainly gathering information, which assist analysis and reporting of cybersecurity analysts.

Here the concepts of Open Source Intelligence (OSINT) and Internet of Things Search Engines (IoTSE) come, which are perfectly suited to collect and analyse publicly available data for investigation purposes (Maltego team, 2020). The popularity of both concepts is increasing in a variety of areas, including but not limited to the detection of open databases (including leaked databases (Bada et al., 2020)). For this reason, i.e. being able to crawl and discover the network of the Internet-connected devices (such as web cameras, databases, industrial automation hardware etc.), they are also known as Search Engines for the Internet of Everything (IoE). The inspection can becarried out at different levels, i.e. (1) system level of only one organization or individual or (2) comprehensive, when overall insight on the state of the art can be gained.

Same as a pentesting by itself, they intend to facilitate identification of whether and at which extent the system tolerates real world-style attack patterns. This information, however, can be used to determine the level of sophistication an attacker needs to successfully compromise the system,countermeasures that could mitigate threats against the system, and defenders' ability to detectattacks and respond appropriately if any are applied to the system, while in opposite case toidentify the need for them and their further usability. The next Section presents the tool, which follows this procedure to identify of open databases by means of IoTSE, demonstrating its potential, covering it from the less often covered viewpoint.

**Search Engine for the Internet of Everything as a tool for detecting vulnerable open data sources**

Search Engine for the IoE-based tool called Shodan- and Binary Edge- based vulnerable open data sources detection tool (ShoBEVODSDT) was originally presented in (Daksevics and Nikiforova, 2021a, 2021b). Therefore, it will not be described in detail, mainly covering its basics. ShoBEVODSDT is a toolfor non-intrusive detection of vulnerable data sources, which is based on the use of Open Source Intelligence (OSINT) tools, more precisely the

Search Engine for the Internet of Everything (also known as Internet of Things Search Engines) – Shodan and Binary Edge. Both are considered to be passive information gathering applications and use Machine Learning (ML) and cybersecurity techniques to scan, acquire and classify public Internet data. In this solution they complement each other, since they outputs tend to slightly differ. The tool conducts a passive assessment, which means that it does not intent to harm the databases - it rather checks for potentially existing bottlenecks or weaknesseswhich, if the attack would take place, could be exposed.

ShoBEVODSDT is designed to search for unprotected databases falling into the list of eight predefined data sources - MongoDB, Redis, Elasticsearch, CouchDB, Cassandra and Memcached, MySQL, PostgreSQL. These sources can be divided in two categories: NoSQL and SQL,while NoSQL can be divided in another three groups: key-value, document-oriented and column-oriented databases. ShoBEVODSDT source code is publicly available on https://github.com/zhmyh/Open-Databases, which allows to contribute to the development of this tool, including its enrichment with additional sources.

The tool overall action can divided into three steps: (1) IP address search (gathering), in scope ofwhich the tool uses BinaryEdge and Shodan libraries to find service IP addresses according to predefined requirements, eliminates duplicates created by the use of these two services and savesthe results in automatically created folders – one per service and country analysed, (2) IP address check aimed at connecting to previously saved IP addresses and registering them depending on the status of the connection, (3) retrieving information from an IP address (parsing), which aims to retrieve the data (if possible) from those databases, to which it has managed to connect. These steps can be called either separately or one-by-one – sequentially.

The data gathered is then classified, by which is meant the examination of the data gathered on the level of their value. Here, previously introduced 6 categories referring to the "value" of the data gathered are used. According to this classification, the data can be matched to: (1) has managed to connect, but failed to gather data, (2) has managed to connect, but the database is empty, (3) has managed to connect by gathering system data or non-sensitive data, (4) has managed to connect and gather sensitive data, (5) compromised database.

Now, let us turn to the results of application of this tool and results demonstrated by the abovementioned databases. Whether MongoDB will be the less secure-by-design? And whether SQL databases will prove to be significantly less vulnerable compared to NoSQL databases?

**Results: whether my database is secure?**

In accordance with Daskevics and Nikiforova (2021), the total number of unprotected or so-called "open" databases, which are available to external actors is less than 2% of the data sources scanned (slightly more than 15 000 IP addresses have been analysed). This percentage is not very high representing 238 IP addresses. However, there are data sources that may pose risks to organizations, where information that can be used for further attacks can be easily obtained. What is more, 12% of identified open data sources havealready been compromised. Thus, let us refer to more detailed results by the database and "value" of data, which is possible to retrieve from it (if any).

Table 1 provides an overall database-wise insight on the results, pointing on whether (1) ShoBEVODSDThas managed to connect to it, (2) has managed to connect, but the database was empty, (3) has managed to connect by gathering system data or non-sensitive information, (4) has managed to connect and gather sensitive data, (5) compromised database found. The first, i.e. "managed to connect", and the latest three categories pointing on the data storage facilities, where the useful for intruders and attackers data and information can be retrieved or have already been retrieved, are critical.

It is shown that CouchDB was the only database, which ShoBEVODSDT has not managed to access Remaining 7 databases were accessed and for every database at least one next "critical" level has been found.

**Table 1. General results by service**

|  | MongoDB | Redis | CouchDB | Memcached | Elasticsearch | Cassandra | MySQL | PostgreSQL |
|---|---|---|---|---|---|---|---|---|
| *Managed to connect* | + | + | - | + | + | + | + | + |
| *Failed to gather data* | - | +/- | - | - | +/- | - | +/- | - |
| *DB is empty* | - | + | - | + | + | + | - | - |
| *System data or non-sensitive* | + | - | - | + | + | + | + | + |
| *Sensitive data* | + | + | - | + | + | - | - | - |
| *Compromised DB* | + | - | - | + | + | - | + | + |

Source: authors

However, this table provides very general results, pointing on the presence or absence of databases falling into one or another category rather than providing quantitative data. In addition, it does not allows to draw conclusion on the level of security neither by the service nor category, i.e. SQL or NoSQL. Although it can be seen that SQL databases indeed cannot be considered as very secured as it could be assumed considering the frequency of debates on the low level of security and high vulnerability of NoSQL databases.

Table 2, however, provides more critical results, presenting a ratio of number of databases falling into particular category related to the total number of databases assessed. They are complemented with the ratio of databases falling into "failed to connect", i.e. thereby proving to be capable not to allow the access to the data stored in it. Ratio instead of numbers should allow to reduce the potential bias of results provided on the basis of analysis of open databases accessible via the Internet, resulted in unequal number of databases found. However, it can still affect them. Therefore, the total number of found IP addresses and databases they refer to are also provided.

**Table 2. Quantitative results by service**

|  | MongoDB | Redis | Memcached | Elasticsearch | Cassandra | MySQL |
|---|---|---|---|---|---|---|
| *Total found* | 177 | 122 | 116 | 86 | 7 | 1347 |
| *Connection successful* | 7.9% | 9.8% | **80%** | **100%** | 14% | 0.14% |
|  |  |  |  |  |  |  |
| *Compromised DB* | **71%** | 0 | 2.2% | **27%** | 0 | 5.3% **33%** |
| *Sensitive data* | 7.1% | **83%** | 24% | 8% | 0 | 0 |
| *Failed to gather data* | 0 | 17% | 0 | 3.5% | 21% | 21% |

Source: authors (partly adapted from (Daskevics and Nikiforova, 2021a))

Memcached and ElasticSearch are the data sources to which ShoBEVODSDT has managed to connect most. To be more precise, ShoBEVODSDT was able to connect to all ElasticSearch instances and 80% of Memcached databases. As regards the sensitivity of the data, Redis is found to be less protected against intrusions and in 83% cases it is possible to obtain sensitive data. It is followed by Memcached, where, however the number of databases from which sensitive data could be gathered is significantly lower, i.e. 24%. Cassandra. MySQL demonstrated the bestresults and there was not possible to gather sensitive data from them.

However, as regards the *compromised databases*, here the results differ and the only facilities, where compromised instances have not been found are Redis and Cassandra, while in the case of MongoDB, 71% databases were already compromised, which is compliant with above mentioned findings made as a result of other studies. However, it is something unexpectedthat the same results have been demonstrated by PostgreSQL, which belongs to SQLdatabases. Even more, the next negative result in this respect is shown by MySQL (33%), which is also SQL database and Elasticsearch (27%). Some compromised databases were also found forMemcached, however their number is significantly lower compared to the above mentioned.

A good point is that the most popular categories are "*has managed to connect by gathering system data or non-sensitive information*" (45%) and "*has managed to connect, but the database is empty*" (21%). However, 18% of inspected data storage facilities had data that could be used by attackers. Even more, as mentioned above, 12% of them have been already compromised.

To sum up, MongoDB is characterized by a high number of cases where databases are compromised (83.3%), while some data sources are not protected from sensitive data gatherings, which can lead in the future to compromising these databases (4.2%), and some databases from which system and non-sensitive data can be gathered, which complies with (Bada et al., 2020) and (Davis, 2018). However, despite its popularity in different studies as the data source characterized by the lowest level of security, unfortunately, most databases inspected are not significantly better. Although PostgreSQL relates to SQL databases, it is still can be characterized by compromised databases, providing the next negative result after MongoDB that are worse than remaining NoSQL databases. In addition, PostgreSQL can be characterized by some databases, from which non-sensitive data or system data can be gathered.

The highest number of open data sources with higher "value" of gathered data were Memcached and ElasticSearch. The only exception is relatively poor result shown by MongoDB for compromised databases and Redis for accessibility to sensitive data. MySQL can be considered as more secure-by-design since in many cases, even if connection to databases was successful, data gathering failed. However, at the same time there were databases from which non-sensitive or system data can be gathered with a few cases of databases from which sensitive data can be gathered and even compromised databases.

According to the statistics provided, it can be assumed that the service with the smallest number of identified vulnerabilities, which may be considered "secure" compared to others, is Cassandra. But in our study CouchDB is the "most secure" because ShoBEVODSDT was unable to connect to any IP address. Although the most "vulnerable" database, according to CVE Details, is MySQL, information-gathering vulnerabilities represent only 2%, which is similar to the data obtained. However, the highest number of the relevant vulnerabilities has been identified for Elasticsearch service (33.3%), which is also in line with our results, where it has managed to connect to all IP addresses and have managed to retrieve information from nearly all instances.

It should also be noted that some results should not be linked to the level of in-built security of the data sources concerned. In some cases, they should also be explained by the database holder's awareness of data security. This could be the reason for surprisingly positive results reported by MySQL, which contradicts the CVE Details. In other words, either MySQL is not characterized as an open database, which, in fact, does not exclude other security-related issues, or the holders of instances in question took care of their security. Otherwise, databases with weaker or any in-built mechanisms are more likely to be vulnerable. What is more, some data sources even do not have authentication mechanism as is the case for Redis, Memcached, while MongoDB and ElasticSearch do not have them enabled by default. MySQL, CouchDB and Cassandra, in turn, require authentication and in most cases show better results when ShoBEVODSDT is used. This makes it possible to argue that even such a primitive and obvious approach as authentication mechanisms leads to a significant reduction of the risk of intrusion, data leakage or even corruption. This does not mean that NoSQL databases should be avoided, but rather points to what Sollins (2019) emphasized in his work that a compromise between IoT big data security and privacy, and innovation should be made and additional security mechanisms should be considered.

## DISCUSSION AND CONCLUSIONS
Companies and individuals, as well as the increasing number of actors representing Cyber-Physical Systems (CPS), Internet of Things and Industrial Internet of Things (IIoT) applications (Oppl, 2022), today create, generate, store, interpret and analyse huge amounts of data. These data and big data, in particular, serve as an input for data-driven processes of decision-making, personalization of services, marketing purposes, crisis management, strategy setting and policy-making and overall performance in both private and public sectors, as well as the whole contemporary society.

The question of the most appropriate data management system is vital. However, when selecting the data storage and management facility, the performance and efficiency of data processing so that they can cope with data that are unmanageable with the traditional / conventional database management systems (Dash et al., 2019; Visvizi & Lytras, 2019) may become the decisive factors.

However, when deciding to use the database, security criteria should also be taken into account in addition to performance, interoperability and cost aspects. Not least important are the four key policies - access control, encryption, data masking and inspection and reporting. In accordance with the principle of security by design and privacy by design, these functions should be provided at all levels: data level (core), connectivity level (middle) and user level (edge) on end devices. The integration of security measures into the database management system is important for the implementation of a consistent cybersecurity concept. It also reduces the necessary efforts to be put at the level of database development, its use / operation and ensuring compliance with legal and internal requirements. This may be perceived by users as a self-evident prerequisite, which would explain why additional security mechanisms are not provided in databases in use, as demonstrate the results of the previous section. It is not least important that the security should be an ongoing priority, constantly checking of both the current security, safety and privacy status of the artefact, and the emerging trends and their appropriateness for further deployment. Today, many emerging technologies are being used in various domains, including security. As a result, more and more promising advanced techniques and solutions based on Big Data (Azeroual and Fabre, 2021), Machine Learning (Azeroual and Nikiforova, 2022)) or distributed ledger technologies (DLT) and blockchains (Abdullah et al., 2022), in particular, appear. They should be constantly examined and included in security policy development, implementation and maintenance, if found appropriate, thereby ensuring sustainable security and safety of the system and its components.

Moreover, it is obvious that the database is not the only asset, which security should be taken into account. A recent study by Verizon (2021) revealed that in 2021 web application and mail servers were the most popular assets affected by incidents, followed by a desktop or laptop, mobile phone (user), a database (server). In terms of patterns, the most popular threats were social engineering, basic web application attacks, system intrusion, miscellaneous errors, privilege misuse, lost and stolen assets, denial of service. This means that the security of IS is very multi-dimensional question, where the questions such as more secure-by-design database management system is probably one of the first questions to be asked. It is even more the case given that, if database security is low, the invention of other even very advanced security mechanisms and users education can fail.

This is the case for data storage and processing facilities, particularly in the context of the widespread databases accessible from outside the organization, which pose the risk to the data holder. These databases can be accessed by external actors, while their data can be both retrieved and even manipulated, including compromising the database. The access to the open databases is simpler than ever before beforehand IoTSEs, such as Shodan, Binary Edge, Censys etc. However, they can also be used for pentesting purposes. In thisChapter one of such tools has been presented. It allows to inspect eight predefined data sources representing both NoSQL and SQL databases.

Although some data storage facilities can be described as sufficiently protected, some of them face serious challenges in this respect. From the "most secure" service viewport, CouchDB has demonstrated very good results in the context of security as the NoSQL database and MySQL as a relational database. However, if the developer needs to use Redis or Memcached, additional security mechanisms and/ or activities should be introduced to protect them. It must be understood, however, that these results cannot be broadly disseminated with regard to the security of the open data storage facility, mostly by demonstrating how many data storage holders were concerned about the security of their data storage facilities, since many data storage facilities have the potential to apply a series of built-in mechanisms.

For the "most unsecure" service, Elasticsearch is characterized by weaker and less frequently used security protection mechanisms. This means that the database holder should be wary of using it. Similar conclusion can be drawn on Memcached (although it contradicts to CVE Details), where the total number of vulnerabilities found was the highest. However, the risk of these vulnerabilities was lower compared to ElasticSearch, so it can be assumed that CVE Details either does not respect such "low-level" weaknesses or have not yet identified them. Here in the future, an in-depth analysis of what CVE Details counts as vulnerability, and further exploration of the correlation with our results, could be carried out.